\documentclass[prx,aps,twocolumn,superscriptaddress]{revtex4-1}

\usepackage[pdftex]{graphicx}
\usepackage{amsbsy,amssymb,amsmath,bm,mathtools}

\usepackage{xspace}
\usepackage{bm}
\usepackage{color}

\usepackage{url}
\usepackage[section]{placeins}
\usepackage[colorlinks=true,linkcolor=blue,citecolor=blue]{hyperref}

\begin{document}


\title{Machine learning predictions for local electronic properties \\ of disordered correlated electron systems}
\author{Yi-Hsuan Liu} 
\affiliation{Department of Physics, National Tsing Hua University, Hsinchu 30013, Taiwan}
\affiliation{Institute of Physics, Academia Sinica, Nankang 11529, Taiwan}

\author{Sheng Zhang} 
\affiliation{Department of Physics, University of Virginia, Charlottesville, Virginia 22904, USA}

\author{Puhan Zhang} 
\affiliation{Department of Physics, University of Virginia, Charlottesville, Virginia 22904, USA}

\author{Ting-Kuo Lee}
\affiliation{Department of Physics, National Tsing Hua University, Hsinchu 30013, Taiwan}
\affiliation{Institute of Physics, Academia Sinica, Nankang 11529, Taiwan}
\affiliation{Department of Physics, National Sun Yat-sen University, Kaohsiun 80424, Taiwan}

\author{Gia-Wei Chern} 
\affiliation{Department of Physics, University of Virginia, Charlottesville, Virginia 22904, USA}

\date{\today}

\begin{abstract}
We present a scalable machine learning (ML) model to predict local electronic properties such as on-site electron number and double occupation for disordered correlated electron systems. Our approach is based on the locality principle, or the nearsightedness nature, of many-electron systems, which means local electronic properties  depend mainly on the immediate environment. A ML model is developed to encode this complex dependence of local quantities on the neighborhood. We demonstrate our approach using the square-lattice Anderson-Hubbard model, which is a paradigmatic system for studying the interplay between Mott transition and Anderson localization. We develop a lattice descriptor based on group-theoretical method to represent the on-site random potentials within a finite region. The resultant feature variables are used as input to a multi-layer fully connected neural network, which is trained from datasets of variational Monte Carlo (VMC) simulations on small systems. We show that the ML predictions agree reasonably well with the VMC data. Our work underscores the promising potential of ML methods for multi-scale modeling of correlated electron systems. 
\end{abstract}

\pacs{Valid PACS appear here}

\maketitle

\section{Introduction}\label{sec:intro}

The growing field of machine learning (ML) is rapidly revolutionizing the scientific research.  In materials science and condensed-matter physics, the ML methods have opened up many research possibilities which are beyond conventional approaches. In particular, the introduction of ML techniques has reinvigorated the field of multi-scale modeling of complex materials.  A fundamental issue in multi-scale simulations is the trade-off between efficiency and accuracy of the numerical methods. In particular, an accurate treatment of complex quantum materials often requires time-consuming calculations, which  significantly limit the accessible system sizes and time scales. 
Recent advances in supervised learning are able to bridge this gap by providing an efficient, yet accurate, model to approximate the outcomes of complicated quantum calculations, thus enabling large-scale simulations. 
A supervised ML model is essentially a complex high-dimensional function with numerous tunable parameters, whose optimal values can be determined from large number of training datasets. Among the various ML models, deep neural networks (NN)~\cite{schmidhuber14,lecun15} represent the most powerful and versatile tools, which, in principle, can approximate any continuous function with arbitrary accuracy~\cite{cybenko89,hornik89,barron93}.  

Perhaps the best example of large-scale modeling enabled by ML is {\em ab initio} molecular dynamics (MD) simulations that are based on ML force-field models~\cite{behler07,bartok10,li15,botu17,li17,smith17,zhang18dp,behler16,deringer19,mcgibbon17,suwa19,mueller20,noe20}. Contrary to classical MD simulations with empirical force fields, the atomic forces in quantum MD are computed by integrating out electrons on-the-fly as the atomic trajectories are generated~\cite{marx09}. Over the past decade, various ML models have been developed to emulate the time-consuming first-principles electronic structure calculations based on, e.g. the density functional theory (DFT). It is worth noting that an ML model here is essentially a complicated classical force-field model, trained from the DFT solutions. ML-based MD simulations thus enjoy the efficiency of classical MD, while maintaining the accuracy of first-principles calculations.

The success of ML method in quantum MD simulations has further motivated similar ML approaches to achieve large-scale dynamical simulations in correlated electron systems~\cite{zhang20,zhang21,zhang21a}. For example, ML methods have been applied to enable large-scale quantum Landau-Lifshitz dynamics simulations of the double-exchange model~\cite{zhang20,zhang21}, which describes itinerant electrons interacting with magnetic moments of localized $d$ electrons. In such simulations, the exchange forces acting on spins are obtained by solving a tight-binding electron Hamiltonian at every time-step, which could be prohibitively expensive for large systems. ML-based exchange-force models is developed to achieve large-scale dynamical simulations of double-exchange systems~\cite{zhang20,zhang21}. In another recent work~\cite{zhang21a}, a multi-layer NN is employed to enable large-scale quantum kinetic Monte Carlo simulations of the Falicov-Kimball model, which is another canonical example of correlated electron systems. 
Furthermore, NN has been employed to learn the Gutzwiller solution of Hubbard-type models~\cite{ma19,suwa19}, thus enabling large-scale MD simulations of Mott metal-insulator transition in atomic liquids~\cite{suwa19}.

It is worth noting that the unprecedented efficiency of ML based multi-scale modeling is due to the linear scalability of electronic structure calculations enabled by ML methods. This is in stark contrast to most conventional approaches which scale rather poorly with the system size. For example, exact diagonalization, which is central to effective single-particle methods including Hartree-Fock, Gutzwiller, and DFT, scales cubically $\mathcal{O}(N^3)$ with the system size $N$. More sophisticated many-body techniques, such as quantum Monte Carlo and configuration interaction, scale even more poorly with increasing system size.

Fundamentally, as first pointed out by W. Kohn, linear-scaling electronic structure methods are possible mainly because of the locality nature or ``nearsightedness" principle~\cite{kohn96,prodan05} of many-electron systems. Indeed, in the pioneering work of Behler and Parrinello~\cite{behler07}, the locality principle was tacitly assumed in their construction of the NN interatomic potential model. The nearsightedness of electronic systems here does not rely on the existence of well localized Wannier-type wave functions, which only exist in large-gap insulators. Instead, Kohn's locality principle mainly refers to observable quantities such as the correlation function  of many-particle systems; the principle is generally a consequence of wave-mechanical destructive interference. It requires the presence of many particles, which need not be interacting. 

Although other linear-scaling methods, notably the kernel polynomial method (KPM)~\cite{weisse06,wang18}, have been developed for electronic structure calculation, they are mostly restricted to solving effective single-electron problem. As a result, they cannot be directly applied to strongly interacting or correlated systems, such as the Hubbard or $t$-$J$ models. Further approximations are required to reduce the many-body Hamiltonian to a single-particle one, which can then be solved by the $\mathcal{O}(N)$ methods such as KPM. On the other hand, assuming nearsightedness for many-electron systems, ML offers a general approach to achieve linear scalability without further approximation. The key is to develop a ML model that can efficiently and accurately emulate the many-body calculations based on a finite local environment.

In this paper, we demonstrate such a scalable ML model for a disordered electron system with Hubbard repulsion. Specifically, we consider the type of on-site potential disorder as described by the Anderson model of localization. Similar on-site disorder can also arise dynamically as in the adiabatic Holstein model. A neural network (NN) model is developed to directly predict local electronic properties such as electron occupation number and double occupancy, based on the disorder configuration in the immediate neighborhood. The NN model is trained from quantum variational Monte Carlo simulations on small lattices. We show that the trained NN model gives accurate predictions on much larger systems of varying Hubbard repulsion and disorder strength.  Our work demonstrates the transferability and scalability of the ML approach to Hubbard-type models, paving the way for their applications to multi-scale modelings of correlated electron systems. 

The remainder of the paper is organized as follows. In Sec.~\ref{sec:vmc}, we present the variational Monte Carlo (VMC) method and details of its implementation to the disordered Hubbard model. Our focus is on the real-space electronic inhomogeneity, and how the local electronic properties depend on the neighborhood disorder configuration. The structure of the NN and the training process are discussed in Sec.~\ref{sec:ml}. A lattice descriptor, based on the group-theoretical method, is developed to incorporate the discrete lattice symmetry into the NN model. Comparisons of the ML predictions versus the VMC results on validation datasets are presented in Sec.~\ref{sec:result}. Moreover, we discuss the application of the ML model to Mott transition of the Anderson-Hubbard model. Finally, a summary and discussion for future work are given in Sec.~\ref{sec:summary}.

\section{Variational Monte Carlo method for Hubbard model} 

\label{sec:vmc}

We consider the following two-dimensional Hubbard model with an on-site disorder, also known as the Anderson-Hubbard model~\cite{ulmke95,dobrosavljevic97,dobrosavljevic03},
\begin{equation}
	\hat H = - t\sum_{\langle ij \rangle,\sigma}  \hat{c}^{\dagger}_{i,\sigma} \hat{c}_{j,\sigma}^{\,} 
	+ \sum_{i} \varepsilon_i \hat{n}^{\,}_{i}
	+\sum_{i} U \hat n_{i\uparrow} \hat n_{i\downarrow}.
\end{equation}
Here $t$ is the nearest-neighbor electron hopping constant, $\hat{c}^\dagger_{i,\sigma}$ is the electron creation operator with spin $\sigma =\uparrow,\downarrow$ at site-$i$,  $\hat{n}_{i, \sigma} \equiv \hat{c}^\dagger_{i, \sigma} \hat{c}^{\;}_{i, \sigma}$ is the corresponding number operator, and $\hat n_i = \hat{n}_{i \uparrow} + \hat{n}_{i \downarrow}$ is the total number operator. The on-site Coulomb repulsion is described by the last term where $U$ is the Hubbard parameter. The second term describes an on-site or potential disorder considered by P. W. Anderson; $\varepsilon_{i}$ denotes the on-site potential, which is a random number drawn uniformly from the interval  $[-W/2,+W/2]$. The strength of the disorder is thus characterized by the parameter $W$.   

It is worth noting that such on-site disorder can also be of dynamical origin, e.g. due to lattice distortions. A canonical example is the Holstein-Hubbard model~\cite{holstein59,zhong92,weber18} in which a scalar dynamical variable $Q_i$ is introduced to describe local lattice distortion of $A_1$ symmetry, such as the breathing mode of oxygen octahedron, associated with site-$i$. The random on-site potential $\varepsilon_i = -g \, Q_i$ comes from the deformation potential coupling between electrons and lattice: $\mathcal{H}_{\rm el\mbox{-}ph} = -g \sum_i \hat{n}_i \, Q_i$, where $g$ is the coupling constant~\cite{holstein59}.  As will be discussed below, the NN model developed for the AH Hamiltonian can also be used to predict the effective forces acting on $Q_i$ in the adiabatic limit.

The Anderson-Hubbard (AH) model is one of the canonical electron systems. In addition to exhibiting rich phase diagrams~\cite{byczuk05,aguiar09,heidarian04,andrade09,lahoud14,patel17,szabo20}, the AH model offers a simple platform to study the interplay between two important mechanisms of metal-insulator transition, namely the Anderson-localization versus correlation-induced Mott transition. The AH model has been extensively studied by a wide variety of numerical methods. Depending on the numerical treatments of the spatial disorder, there are two types of approaches to this problem: the self-consistent theories and the real-space methods. The most representative example of the former approach is the generalization of dynamical mean field theory (DMFT)~\cite{georges96} to include the disorder effects. While several self-consistent theories have been developed for disordered electronic systems, the typical medium theory (TMT)~\cite{dobrosavljevic03b} proves successful to capture the Anderson localization phenomena and can be readily combined with DMFT~\cite{byczuk05,aguiar09}. As such self-consistent methods are free of finite-size effect, they provide an overall theoretical picture of the AH model in the thermodynamic limit. 

The non-magnetic phase diagram of AH model obtained from the TMT-DMFT method includes three distinct phases: a correlated metallic phase, a Mott insulating phase, and an Anderson insulating phase~\cite{byczuk05,aguiar09}. Importantly, the two insulating phases of the AH model have very different characters. The Mott insulator results from the strong correlation effect which prohibits electrons from hopping to the neighboring sites. On the other hand, strong disorder weakens the constructive interference that allows an electron wave packet to propagate coherently in a periodic potential, leading to the Anderson insulator. TMT-DMFT calculation shows that these two insulating phases are continuously connected~\cite{byczuk05,aguiar09}.

On the other hand, the real-space approach, although limited by finite-size effect, can better describe the spatial fluctuations and correlations of the inhomogeneous electronic state due to the disorder, especially for low-dimensional systems. In this real-space approach, the AH Hamiltonian for a particular disorder configuration on a finite lattice is solved by many-body techniques ranging from unrestricted Hartree-Fock~\cite{tusch93,fazileh06,heidarian04,shinaoka09} and Gutzwiller~\cite{andrade09,andrade10} mean-field type theories to small-cluster exact diagonalization~\cite{kotlyar01,srinivasan03,chiesa08}, inhomogeneous or statistical DMFT~\cite{dobrosavljevic97,tanaskovic03,villagran20,semmler11}, and quantum determinent Monte Carlo~\cite{sandvik93,sandvik94,otsuka98,ulmke97} as well as variational Monte Carlo simulations~\cite{pezzoli09,pezzoli10,farhoodfar09,farhoodfar11}. The calculation results are then averaged over different disorder realizations. 
In particular, extensive large-scale simulations based on the Gutzwiller/slave-boson methods showed that the strong spatial inhomogeneity gives rise to an electronic Griffiths phase that precedes the metal-insulator transition~\cite{andrade09}. 


In this work, we are interested in the locality of electronic properties and collective behaviors such as the double-occupancy. Specifically, our goal is to develop a NN model that can accurately predict on-site quantities based on disorder configuration within a finite neighborhood. To capture these spatial inhomogeneity, we employ the real-space variational Monte Carlo (VMC) method to solve the square-lattice AH model. As we are interested mainly in the competition between localization effect and electron correlation, we restrict ourselves to the paramagnetic phases to avoid complications due to long-range magnetic order.  We note that a similar ML approach to disordered Hubbard models has been demonstrated based on dataset from the real-space Gutzwiller/slave-boson method~\cite{ma19}. However, the spatial correlation between local electronic properties is ignored due to the Gutzwiller approximation. The VMC method, on the other hand, can properly take into account these spatial correlations, which are important to test the locality of the ML models.

Next we outline the VMC method for the square-lattice AH model. Following the previous works developed for Hubbard-type models, we consider a variational wave function obtained by applying  a Gutzwiller factor $\hat{\mathcal{G}}$ and a Jastrow factor $\hat{\mathcal{J}}$ to a Slater determinant~$|\Phi_0\rangle$~\cite{pezzoli09,pezzoli10,capello05}:
\begin{equation}
	|\Psi\rangle = \hat{\mathcal{G}} \hat{\mathcal{J}} |\Phi_0\rangle.
\end{equation}
The uncorrelated Slater determinant state $|\Phi_0\rangle$ is computed from the eigenstates of the following quadratic Hamiltonian
\begin{equation}
	\hat H_{\rm MF} = -t \sum_{ij,\sigma} \hat{c}^{\dagger}_{i,\sigma} \hat{c}^{\,}_{j,\sigma} 
	+ \sum_{i,\sigma}  \tilde{\varepsilon}_{i, \sigma} \, \hat{n}^{\,}_{i,\sigma}, 
\end{equation}
which can be viewed as a mean-field approximation to the AH model. The on-site energies $\tilde \varepsilon_{i, \sigma}$ are part of the variational parameters. As mentioned above, we focus on the paramagnetic phases and assume spin-independent local energies $\tilde \varepsilon_{i, \uparrow} = \tilde \varepsilon_{i, \downarrow} = \tilde \varepsilon_i$. To account for the crucial on-site electron correlation, the Gutzwiller correlator is introduced~\cite{gutzwiller63,gutzwiller65,capello05}
\begin{equation}
	\hat{\mathcal{G}} = \prod_i \bigl[ 1 - (1- g_i ) \, \hat n_{i\uparrow} \hat{n}_{i\downarrow} \bigr] \equiv \prod_i \hat{\mathcal{G}}_i,
\end{equation}
where $g_i$ are another set of variational parameters that control the on-site double occupancy; as $g_i \to 0$, double-occupied states are completely projected out. 
The long-range Jastrow factor is defined as~\cite{capello05}
\begin{equation}
	\hat{\mathcal{J}} = \exp\biggl[ \frac{-1}{2}\sum_{ij} v_{ij} (\hat n_i-1) (\hat n_j - 1 ) \biggr],
\end{equation}
The Jastrow operator, parameterized by another set of variational parameters $v_{ij}$, introduces correlation of charge fluctuations $\delta n_i = n_i - 1$ at different sites.  For an arbitrary inhomogeneous state, the parameter $v_{ij}$ in principle depends on both site-$i$ and $j$, giving rise to a total of $N^2$ parameters to be optimized. In order to make the numerical calculation feasible~\cite{pezzoli09,pezzoli10}, we assume translational invariance for these parameters, i.e. $v_{ij} = v(| \mathbf r_i - \mathbf r_j|)$, and consider $v_{ij}$ for different pairs up to the 8th nearest-neighbors.

The optimal variational parameters $\tilde \varepsilon_i, g_i$, and $v_{ij}$ are obtained by minimizing the variational energy $E_{\rm var} = \langle \Psi | \hat{\mathcal{H}} | \Psi \rangle / \langle \Psi | \Psi \rangle$ using the stochastic reconfiguration method~\cite{sorella05,becca17}. The evaluation of the various expectation values computed from $|\Psi\rangle$ is computed based on Monte Carlo simulations. In the following we apply the VMC methods to study the ground state of the AH model with various disorder strength $W = 6t, 10t, 14t, 18t$ and Hubbard parameter $U = 4t, 8t, 10t, 12t, 16t$. We focus on the case of half-filling where the number of electrons $N_e = N$. Periodic and anti-periodic boundary conditions along $x$ and $y$ directions, respectively, are used. 
Depending on the convergence, 500--2000 iterations of the stochastic reconfiguration were used to optimize the variational parameters. For each iteration, the various expectation values were obtained from approximately $10^5$ Monte~Carlo samplings. For each $U$ and $W$ combination, 50 independent realizations were used to generate the training datasets.  


\begin{figure}
\includegraphics[width=0.99\columnwidth]{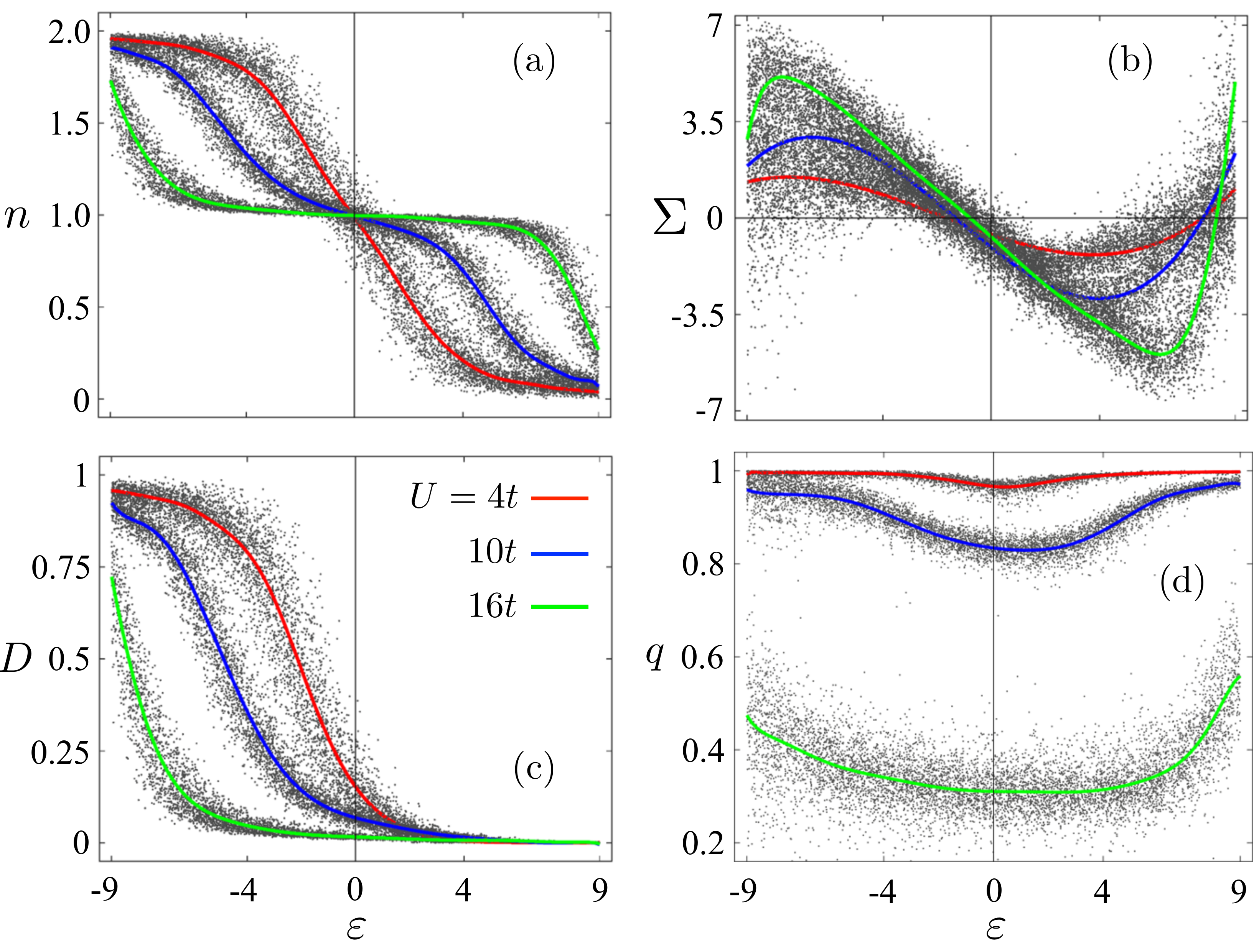}
\caption{Summary of the VMC solution for the AH model on a $16 \times 16$ square lattice. The four panels show the scatter diagram of  (a) on-site electron number $n_i$, (b) variational local self-energy $\Sigma_i$, (c) double occupancy $D_i $, and (d) effective quasi-particle weight $q_i$ versus the random on-site potential $\varepsilon_i$. The data points were obtained from calculations with random strength $W/t = 6, 10, 14, 18$ and three different $U = 4t, 10t$, and $16t$. The overall dependence of the various local quantities on the on-site potential is highlighted by the three colored curves obtained using polynomial regression with up to 16th-order polynomials. The red, blue, and green curves correspond to $U = 4t$, $10t$, and $16t$, respectively.
\label{fig:vmcdata} 
}
\end{figure}

The results of the VMC simulation are summarized in Fig.~\ref{fig:vmcdata}, which shows the scatter plots of various local quantities versus on-site potentials $\varepsilon_i$ for three different Hubbard $U$. Here we are interested in the following local quantities: (a) the local electron filling fraction $f_i$, (b) the electron on-site self-energy $\Sigma_i$,  (c) the double-occupancy $D_i$, and (d) the local quasi-particle weight $q_i$. Each point in the scatter plots corresponds to the data of a lattice site from the VMC solution of a given disorder realization.  
The local electron number is defined as
\begin{eqnarray}
	n_i = \langle \hat{n}_i \rangle = \langle \hat{n}_{i, \uparrow} \rangle + \langle \hat{n}_{i, \downarrow} \rangle.
\end{eqnarray}
for paramagnetic phases, we have $ \langle \hat{n}_{i, \uparrow} \rangle = \langle \hat{n}_{i, \downarrow} \rangle$.
As shown in Fig.~\ref{fig:vmcdata}(a), this local electron density is reduced with increasing on-site potential, which is expected. At large $U = 16t$, the electron number develops a plateau at $n_i = 1$ for small on-site energies $|\varepsilon_i| \lesssim 6t$. The plateau thus represents lattice sites with localized electrons, where the corresponding double occupancy also tends to zero.  The electronic properties outside the plateau is dominated by the strong local potential, leading to either almost filled or empty sites. For systems with large $W \gtrsim U$, the interplay between electron correlation and disorder thus gives rise to a spatially very inhomogeneous states with coexisting Mott regions and Anderson insulator, consistent with the two-fluid behavior of the Mott-Anderson insulator~\cite{andrade09,aguiar09,villagran20}

Another quantity of interest is the local self-energy, which is defined as the difference between the renormalized and bare on-site potentials:
\begin{eqnarray}
	\Sigma_i = \tilde\varepsilon_i - \varepsilon_i.
\end{eqnarray}
As shown in Fig.~\ref{fig:vmcdata}(b), this self-energy clearly anti-correlates with the random on-site potential $\varepsilon_i$, meaning that the renormalization due to $\Sigma_i$ is such that the effective potential $\tilde \varepsilon_i$ tends to vanish. Indeed, screening of impurity potential by the electron gas has been demonstrated even in the weak interaction regime. As demonstrated in both real-space DMFT and slave-boson studies on the 2D AH model~\cite{andrade09,andrade10,tanaskovic03}, the disorder screening persists also in the strong correlation regime, albeit with a rather different nature.  In particular, this interplay leads to the enhancement of metallicity in an intermediate regime where the interactions and the disorder are of comparable magnitude.

To detect the correlation-induced electron localization, we compute the local double occupancy from VMC:
\begin{eqnarray}
	D_i  = \langle \hat{n}_{i, \uparrow} \hat{n}_{i, \downarrow} \rangle,
\end{eqnarray}
As expected, the average double occupation is reduced with increasing Hubbard $U$; see Fig.~\ref{fig:vmcdata}(c). It should be noted that the large $D_i$ persists even at large $U$ is due to the deep on-site potential which traps two electrons of opposite spins.  On the other hand, the small value of double-occupancy at large positive $\epsilon$ is a result of empty site, instead of strong correlation.  To properly distinguish these two scenarios, we also compute an effective local quasi-particle weight defined as
\begin{eqnarray}
	q_i = \frac{4 g_i}{(1 + g_i)^2},
\end{eqnarray}
where $g_i$ is obtained from the VMC optimization. For homogeneous electron liquids, the quantity $q$ characterizes the discontinuity of the momentum distribution function at the Fermi surface in the Gutzwiller approximation~\cite{gutzwiller63,gutzwiller65}. And for inhomogeneous systems, $q_i$ plays the role of renormalizing the electron hopping. We emphasize that $q_i$ is not an exact definition, but is meant to be a qualitative indicator of the local quasi-particle weight. As shown in Fig.~\ref{fig:vmcdata}(d), overall the quasi-particle weights decrease with increasing Hubbard $U$. On the other hand, stronger disorder, e.g. sites with $|\varepsilon_i| \lesssim U$ preserves the itinerant nature of electrons.

\begin{figure*}
\includegraphics[width=1.98\columnwidth]{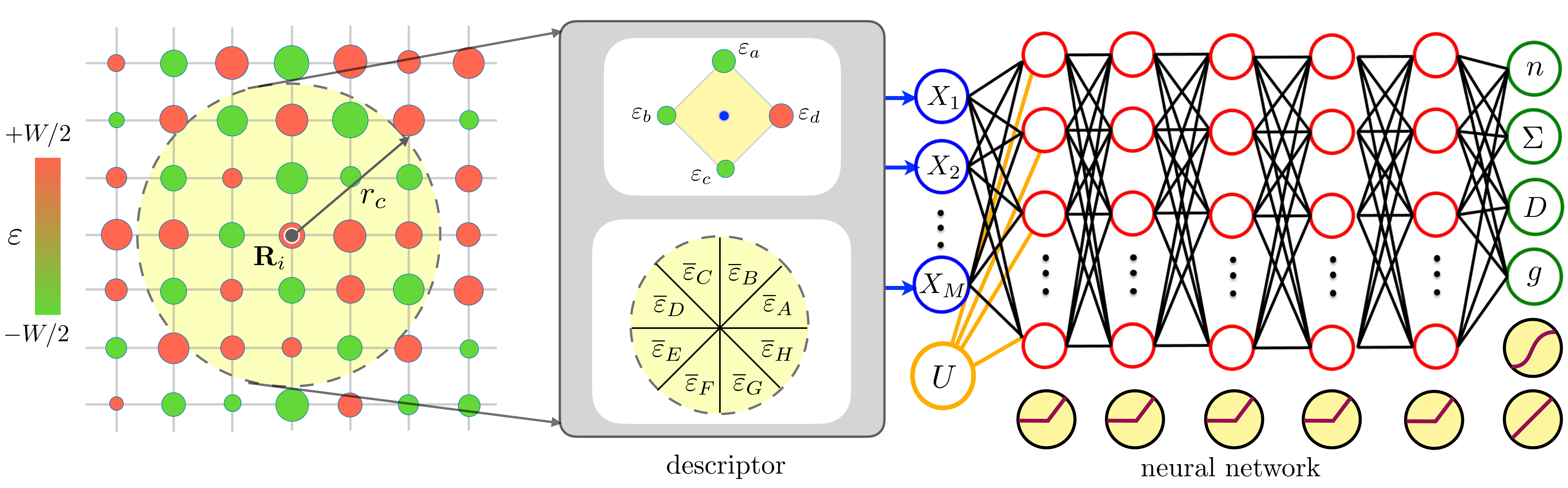}
\caption{Schematic diagram of the ML model for the vector function $\boldsymbol{\mathcal{F}}(\mathcal{C}_i; U)$ defined in Eq.~(\ref{eq:vector-fx}) for the Anderson-Hubbard model. The input of the ML model is the disorder configuration $\mathcal{C}_i$ in the neighborhood of a given site-$i$ up to a cutoff radius~$r_c$.  The output are local electron properties including on-site electron number $n_i$, self-energy $\Sigma_i$, double-occupancy $D_i$, and Gutzwiller parameter $g_i$. There are two central components of the ML model: the descriptor and the learning model, which is based on a multi-layer neural network. The ReLU activation function is used in the five hidden-feature extraction layers with $256 \times 128 \times 64 \times 32$ nodes. Note the special input node corresponding to the Hubbard parameter $U$.
\label{fig:ml-model} 
}
\end{figure*}

The results summarized in Fig.~\ref{fig:vmcdata} also indicate continuous trend with respect to increasing disorder for a fixed Hubbard parameter. We note that these scatter plots include VMC data from various disorder strengths $W/t = 6$, 10, 14, and 18. As a result, some of the data points correspond to the Mott-dominated insulating phase, i.e. those $W < U$, while others belong to the Anderson insulator ($W > U$).  Nonetheless, the collection of all data points clearly exhibit an overall trend as a function of the on-site potential $\varepsilon$ as highlighted by the smooth curves in Fig.~\ref{fig:vmcdata}, which are obtained using polynomial regression up to 18th order. This continuous dependence on the on-site potential $\varepsilon_i$ thus provides the zeroth order prediction for the various local quantities. Deviations from this smooth curve, as clearly indicated by the scattered points in the figure, thus can be viewed as due to the influence of neighboring random potentials. The effects of the neighborhood disorder are accounted for by the ML model to be described below.

\section{Machine learning model}

\label{sec:ml}

We next describe the framework of a scalable ML model for predicting the local electronic properties of disordered Hubbard systems. Our approach is similar to the ML modeling of structure-property relationships in materials science, which play an increasingly important role in accelerated materials discovery. A particularly important application, as mentioned in Sec.~\ref{sec:intro}, is the ML modeling of force-field for {\em ab initio} MD simulations. The central idea is to develop a ML model which can accurately predict the force acting on individual atoms and other local properties based on the immediate chemical environment. 

A widely used scheme, first pioneered by Behler and Parrinello~\cite{behler07}, focuses on local energies. First, the total energy of the system is partitioned into local contributions, $E = \sum_i \epsilon_i$, where $\epsilon_i$ is called the atomic energy associated with the $i$-th atom. Next, based on locality principle, the atomic energy $\epsilon_i$ is assumed to depend on the atomic configuration $\mathcal{C}_i$ in the neighborhood of atom-$i$. Specifically, this local chemical environment is given by $\mathcal{C}_i =\{ (Z_j, \mathbf R_j) \, \big| \, |\mathbf R_j - \mathbf R_i | < r_{c} \}$, where $Z_j$ is the atomic number of atom-$j$ at position $\mathbf R_j$, and $r_c$ is a cutoff radius (soft cutoff is often used).  Finally, the complex dependence of atomic energy on the local environment, $\epsilon_i = f( \mathcal{C}_i)$, is to be approximated by an ML model, which is trained from electronic structure calculations such as DFT on small systems. 

The atomic forces in this scheme is obtained from the derivatives of the total energy: $\mathbf F_i = - \partial E / \partial \mathbf R_i$. It is worth noting that, instead of direct prediction of atomic forces, which are vectors, the Behler--Parrinello type scheme focuses on the local atomic energy. The fact that the atomic energy, as a scalar, is readily invariant under transformations such as rotations and translations of the system makes it easier to incorporate the symmetry properties into the ML model. Moreover, as forces are derived from an effective energy, this approach also ensures a conservative force field, which is important for quantum MD under Born-Oppenheimer approximation~\cite{marx09}. Most importantly, this ML method is both transferrable and scalable as exactly the same ML model can be used for much larger systems. We note in passing that a similar approach has recently been developed for generalized force fields in condensed matter systems~\cite{zhang22}. 

Here we adapt this ML approach to develop a neural-network (NN) model for the prediction of local electronic properties of the AH model. 
Again, based on the nearsightedness of many-electron systems~\cite{kohn96,prodan05}, we assume local electronic properties such as  electron density and double-occupancy at site-$i$ only depends on the neighborhood disorder configuration~$\mathcal{C}_i$. Explicitly, it is defined~as
\begin{eqnarray}
	\mathcal{C}_i = \left\{ \varepsilon_j \, \big| \, |\mathbf R_j - \mathbf R_i | < r_c \right\},
\end{eqnarray}
where $r_c$ is a predefined cutoff radius. The complex dependence of local electronic properties on the neighborhood is represented by a vector function:
\begin{eqnarray}
	\label{eq:vector-fx}
	\boldsymbol{\mathcal{Q}}_i = \bigl( n_i, \Sigma_i, D_i, g_i, \cdots \bigr) = \boldsymbol{\mathcal{F}}\bigl(\mathcal{C}_i;  \,U \bigr).
\end{eqnarray}
For convenience, here we arrange the local quantities associated with site-$i$ into a vector or array $\bm{\mathcal{Q}}_i$. The definitions of these quantities are discussed in Sec.~\ref{sec:vmc}. Although the approach discussed here can be straightforwardly generalized to include more local properties, here we mainly concern the four quantities shown in Fig.~\ref{fig:vmcdata}. Moreover, we have explicitly included the dependence on the Hubbard parameter $U$. By setting the nearest-hopping constant $t = 1$, which effective serves as the unit for energy, the vector function $\boldsymbol{\mathcal{F}}(\cdot)$ is universal for the AH model of a given electron filling fraction $n = N_e / N$. 
A ML model, shown in Fig.~\ref{fig:ml-model}, is developed to approximate this universal function for the case of half-filling. The input of the model, which is the disorder configuration $\mathcal{C}_i$ in the neighborhood of site-$i$, is first transformed into a set of feature variables $\{ x_1, x_2, \cdots, x_M \}$ called the descriptor. These feature variables, along with the Hubbard parameter $U$, are then fed into the neural network which produces an array of the local quantities $\boldsymbol{\mathcal{Q}}_i$ at the output. 
Details of these two central components, namely the descriptor and the neural network, are discussed in the following.

\subsection{Lattice Descriptor}

\label{sec:descriptor}

It is worth noting that, despite the powerful approximation capability of NNs, symmetries of the electron Hamiltonian are not automatically included in the ML model. One approach to incorporate the required symmetry into ML model is through the construction of a proper representation of the local environment, which is then used as input to the learning model. A good representation is {\em invariant} with respect to transformations of the symmetry group of the system. This crucial step of the ML model, namely the construction of the proper representation, is often referred to as feature engineering and the resultant feature variables, also called the generalized coordinates, are termed a descriptor~\cite{ghiringhelli15,bartok13,himanen20}. 

Descriptor is also crucial to the ML interatomic potentials for quantum MD simulations.  a proper description of the chemical environment should respect the fundamental symmetries of interatomic interactions, which are invariant under translational, rotational, and permutational transformations.  Over the past decade, a number of descriptors have been proposed together with the learning models based on them~\cite{rupp12,behler07,behler11,bartok10,bartok13,zhang18dp,ghiringhelli15,bartok13,himanen20,steinhardt83,behler11,shapeev16,drautz19,hansen15,faber15,huo18}. A popular atomic descriptor used in many ML models is the atom-centered symmetry functions (ACSFs) built from the two-body (relative distances) and three-body (relative angles) invariants of the atomic configurations~\cite{behler07,behler11}. The group-theoretical method, on the other hand, offers a more controlled approach to the construction of atomic representation based on the power-spectrum and bispectrum coefficients~\cite{bartok10,bartok13}. It is worth noting that the research of atomic descriptor is an active ongoing field.

A general theory and several specific implementations of descriptors in condensed matter systems, especially for lattice models, have recently been presented in Ref.~\cite{zhang22}. In particular, the group-theoretical bispectrum method was generalized to systematically generate feature variables that are invariant under symmetry operations of the on-site point group~\cite{zhang22,ma19}. Here we apply this method to develop a descriptor of the AH model. To this end, we first note that the on-site potentials in the neighborhood $\mathcal{C}_i$ form a high-dimensional reducible representation of the site-symmetry group, which in the case of square lattice is equivalent to the point group $D_4$. The first step of finding invariants under site-symmetry is to decompose the neighborhood $\mathcal{C}_i$ into irreducible representations (IRs) of the symmetry group.

\begin{figure}
\includegraphics[width=0.99\columnwidth]{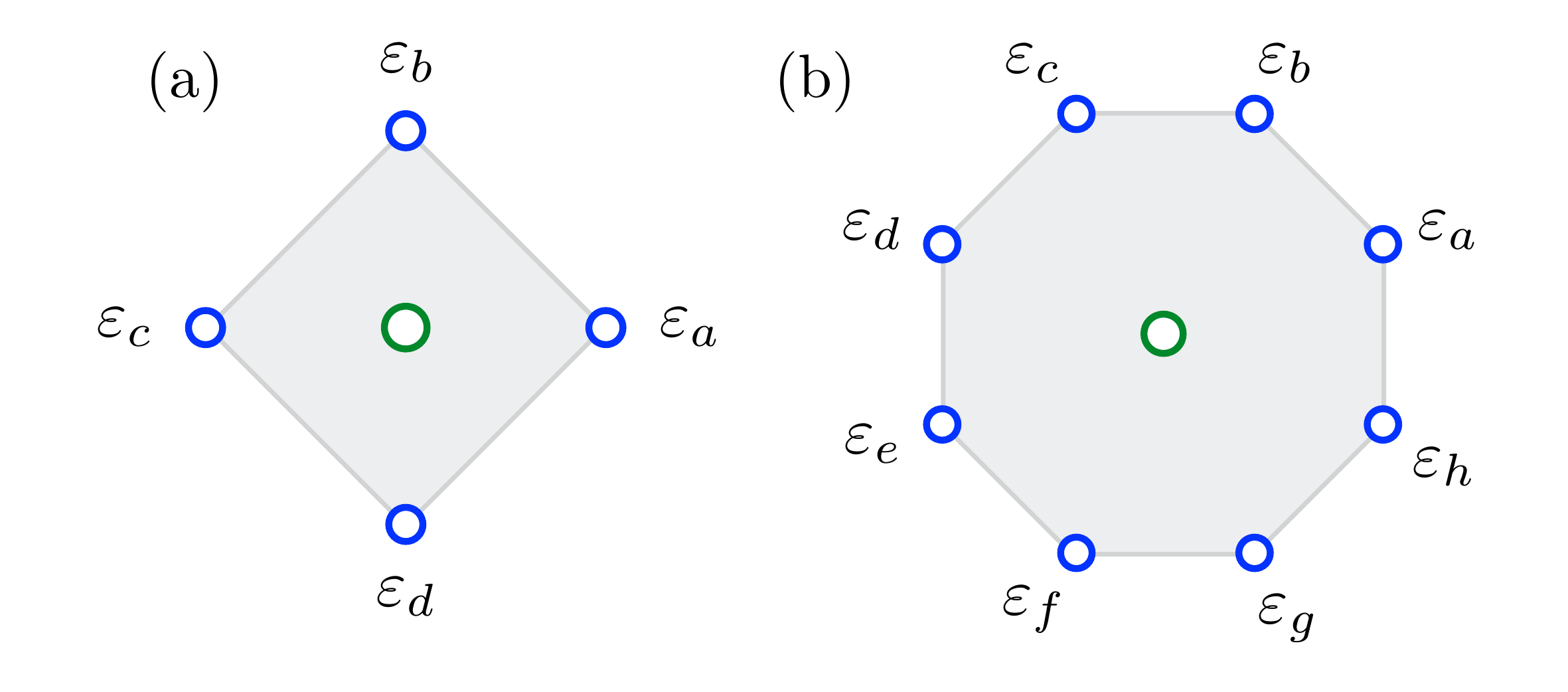}
\caption{Schematic diagrams of (a) 4-sites and (b) 8-sites blocks of neighbor sites.
\label{fig:blocks} 
}
\end{figure}

While this decomposition can be done following standard procedure in group theory, the calculation here can be greatly simplified by noting that neighboring sites-$j$ with the same distance $R_{ij} = |\mathbf R_j - \mathbf R_i|$ from the center site-$i$ form a closed representation of the point group. In the case of $D_4$, the size of these invariant neighbor blocks is either 4 or 8; see Fig.~\ref{fig:blocks}. The 4-sites block can be decomposed as: $4 = 1A_1 \oplus 1B_1 \oplus 1E$. The expansion coefficients of each IR are 
\begin{eqnarray}
	& & f_{A_1} = \varepsilon_a + \varepsilon_b + \varepsilon_c + \varepsilon_d, \nonumber \\
	& & f_{B_1} = \varepsilon_a - \varepsilon_b + \varepsilon_c - \varepsilon_d, \\
	& & \bm f_E = (\varepsilon_a - \varepsilon_c, \, \varepsilon_b - \varepsilon_d). \nonumber
\end{eqnarray}
The decomposition of the 8-sites block is: $8 = 1A_1 + 1B_1 + 1A_2 + 1B_2 + 2 E$, with the following coefficients for each IR 
\begin{eqnarray}
	& & f_{A_{1}}  = \varepsilon_a + \varepsilon_b + \varepsilon_c + \varepsilon_d + \varepsilon_e + \varepsilon_f + \varepsilon_g + \varepsilon_h, \nonumber \\
	& & f_{A_{2}}  = \varepsilon_a - \varepsilon_b + \varepsilon_c - \varepsilon_d + \varepsilon_e - \varepsilon_f + \varepsilon_g - \varepsilon_h, \nonumber \\
	& & f_{B_{1}}  = \varepsilon_a - \varepsilon_b - \varepsilon_c + \varepsilon_d + \varepsilon_e - \varepsilon_f - \varepsilon_g + \varepsilon_h, \nonumber \\
	& & f_{B_{2}}  = \varepsilon_a + \varepsilon_b - \varepsilon_c - \varepsilon_d + \varepsilon_e + \varepsilon_f - \varepsilon_g - \varepsilon_h, \nonumber \\
	& & \bm f_{(E, 1)}  = (\varepsilon_a + \varepsilon_b - \varepsilon_e - \varepsilon_f,\  -\varepsilon_c - \varepsilon_d + \varepsilon_g + \varepsilon_h), \nonumber \\
	& & \bm f_{(E, 2)}  = (\varepsilon_c - \varepsilon_d - \varepsilon_g + \varepsilon_h,\  \varepsilon_a - \varepsilon_b - \varepsilon_e + \varepsilon_f).
\end{eqnarray}
As the neighborhood $\mathcal{C}_i$ contains several such invariant blocks, we expect same IRs appear multiple times in the overall decomposition of $\mathcal{C}_i$. In the following, we label each IR in the decomposition of $\mathcal{C}_i$ as $\Gamma = (\mathtt{T}, r)$, where $\mathtt{T} = A_1, A_2, \cdots$ denotes the symmetry type of the IR, and $r$ indicates different occurrence of the same IR. 
For convenience, we arrange the expansion coefficients of an IR $\Gamma$ into a vector $\bm f_\Gamma = (f_{\Gamma, 1}, f_{\Gamma, 2}, \cdots, f_{\Gamma, n_\Gamma} )$, where $n_\Gamma$ is the dimension of $\Gamma$.

The power spectrum of the representation are given by the amplitudes of the IR coefficients
\begin{eqnarray}
	p_{\Gamma} = \bigl| \bm f_\Gamma \bigr|^2.
\end{eqnarray}
Since the power spectrum coefficients are obviously invariant under symmetry transformations, they can be used as feature variables for the ML models. However, a descriptor composed only of power spectrum is incomplete since the relative phases between different IRs are ignored. This also means that descriptor contains spurious symmetries as the transformation of each IR is independent of each other without the phase information.  A complete set of feature variables can be obtained from the bispectrum coefficients $b_{\Gamma_1, \Gamma_2, \Gamma_3}$, which are triple products of the expansion coefficients $\bm f_{\Gamma_{1, 2, 3}}$ based on the Clebsch-Gordan coefficients of the point group. Intuitively, they can be viewed as the analog of scalar triple product of 3-dimensional vectors. Not only are the bispectrum coefficients invariant under symmetry transformations, they can also be used to faithfully reconstruct the original disorder configuration [...].

However, a descriptor based on all the bispectrum coefficients is in fact over-complete as many of them are redundant. Moreover, since the dimension of most IRs of point groups is rather small, the number of bispectrum coefficients is often a very large number, which makes the implementation infeasible. Instead, here we employ the method of reference IR discussed in Ref.~\cite{zhang22} to retain the phase information. The central idea is to first construct an 8-dimensional representation of the neighborhood $\mathcal{C}_i$ based on average of on-site potentials over symmetry-related finite regions. As shown in Fig.~\ref{fig:ml-model}, an example is given by $(\overline{\varepsilon}_A, \overline{\varepsilon}_B, \cdots, \overline{\varepsilon}_H)$ where each $\overline{\varepsilon}_K$ is given by the average of all on-site $\varepsilon_j$ within wedge-$K$. 

The decomposition of this 8-dimensional representation $\overline{\varepsilon}_K$ then gives coefficients $ f_{A_1}^*$, $ f^*_{A_2}$, $\cdots$, $\bm f^*_E$ for each symmetry type. These coefficients $\bm f^*_{\mathtt{T}}$ are termed the reference IR coefficients. For each IR, an effective phase can be defined by the following inner product 
\begin{eqnarray}
	\eta_\Gamma = \bigl( \bm f_\Gamma \cdot \bm f^*_{\mathtt{T}_\Gamma} \bigr) / \bigl| \bm f_\Gamma \bigr| \bigl| \bm f^*_{\mathtt{T}_\Gamma} \bigr|, 
\end{eqnarray}
where $\mathtt{T}_\Gamma$ is the symmetry type of IR~$\Gamma$. The phase $\eta_\Gamma$, which is an inner product of two IR coefficients, is naturally invariant with respect to symmetry operations. More importantly, by including $\eta_\Gamma$ in the descriptor, the relative phases between different $\bm f_\Gamma$ are now be inferred through the intermediate reference IR coefficients.

\subsection{Neural Network}

\label{sec:nn}

The various steps of the descriptor discussed above can be summarized as: $\mathcal{C}_i \, \rightarrow \, \bm f_\Gamma \,  \rightarrow \, (p_\Gamma, \eta_\Gamma)$. Crucially, assuming the various local quantities $\bm{\mathcal{Q}}_i$ of interest depend on the neighborhood through these feature variables, 
\begin{eqnarray}
	\bm{\mathcal{Q}}_i = \bm{\mathcal{F}}\bigl( \{ p_\Gamma, \eta_\Gamma \}_i ; U \bigr).
\end{eqnarray}
the resultant ML model is ensured to preserve the site symmetry of the AH Hamiltonian. As discussed above, this vector function $\bm{\mathcal{F}}(\cdot)$ is to be implemented using a NN. The basic unit of a NN is a perceptron or artificial neuron. And a NN is essentially a set of nested linear regression functions with non-linear activation performed by the neurons; see Fig.~\ref{fig:ml-model}. For a neuron with $m$ input signals, arranged in a vector $\bm x = (x_1, x_2, \cdots, x_m)$, its output is given by $y = \sigma( \bm w \cdot \bm x + b)$, where $\sigma(\cdot)$ is a non-linear activation function, $\bm w = (w_1, w_2, \cdots, w_m)$ specifies weights for each input, and $b$ denotes a bias. Here each signal $x_k$ is the output of a neuron from previous layer. In a NN, each line is associated with a weight, while each node (neuron) is assigned a bias. These weight and bias variables are parameters to be optimized through training processes. 

We design a fully-connected NN with four hidden layers consisting of $\mathsf{N} = 256\times128\times 64 \times 32$ rectified linear units (ReLU) neurons, i.e. with an activation function $\sigma(x) = {\rm max}(0, x)$. The input layer, specified by a vector $\bm X = ( X_1, X_2, \cdots, X_M )$, are given by the standard-scalar transformation, i.e. by removing the mean and scaling to unit variance, of the power spectrum coefficients $p_\Gamma$ and the relative phases $\eta_\Gamma$. In addition, the Hubbard parameter $U$ is also used as an input to the NN; see Fig.~\ref{fig:ml-model}. The neurons in the hidden layers perform a series of nonlinear transformations described above on the input data. The outcome is fed forward to be processed by the output neuron with sigmoid activation function for $n$, $D$, and $g$ (whose domain is  $[0,1]$) and linear activation function for~$\Sigma$.
The mean squared error (MSE) is used as the loss function:
\begin{eqnarray}
	& & L = \frac{1}{\mathcal{N}} \sum_{k=1}^{\mathcal{N}} \Big( W_n \bigl| n_k - \hat{n}_k \bigr|^2 +W_\Sigma \bigl| \Sigma_k - \hat{\Sigma}_k \bigr|^2  \\
	& & \qquad \qquad \qquad+ W_D \bigl| D_k - \hat{D}_k \bigr|^2 + W_g \bigl| g_k - \hat{g}_k \bigr|^2 +  \cdots \Big), \nonumber
\end{eqnarray}
where $\mathcal{N}$ is the number of training data, symbols with a hat refer to the predicted values, the various $W$ denote the weights of each output.  The method of batch normalization is used to avoid overfitting, with a minimum batch size of~32. We use randomly mixed $36\times256\times4\times5=184320$ data samples as the training set. The Adam optimizer with learning rate of 0.001 is applied for training process. Once the training process is successful, the trained neural network can rapidly predict the  $4\times256\times4\times5=20480$ test data samples. Equal weights for the output are used in the current model.

It is worth noting that, instead of developing an independent NN for each of the local quantities, here adopt a multi-task ML framework~\cite{caruana97,collobert08}. As shown in Fig.~\ref{fig:ml-model}, our approach is to build one NN that can simultaneously and consistently predict different local electronic properties. This common NN is trained via a loss function $L$ that includes MSE from all four local quantities introduced in Eq.~(\ref{eq:vector-fx}).  Such multitask learning approach allows inductive bias to be acquired via the training signals for related additional tasks drawn from the same domain~\cite{caruana97}. The benefit of multitask learning is the additional constraints due to the interdependences between the multiple outputs; what is learned for each task can help other tasks to be learned better.

\section{results and discussion}

\label{sec:result}

\begin{figure}
\includegraphics[width=0.99\columnwidth]{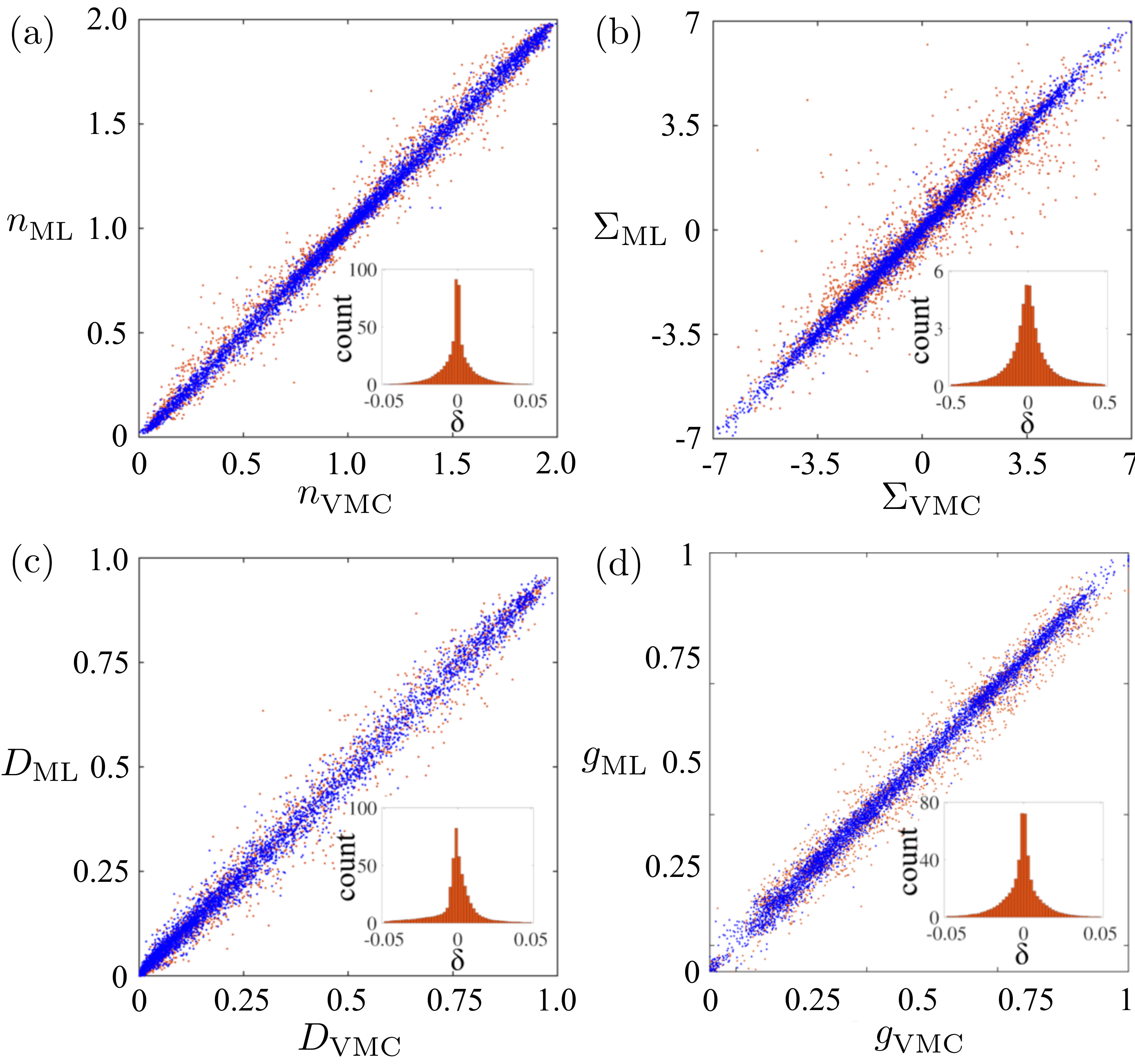}
\caption{Comparison of the ML predictions with references obtained from the VMC simulations. (a) local electron number $n$, (b) on-site electron self-energy $\Sigma$, (c) local double-occupancy $D$, and (d) the Gutzwiller variational parameter~$g$. The blue and orange data points denote predictions for training and test data sets, respectively. The insets show the normalized count of the error $\delta= A_{\rm ML} - A_{\rm VMC}$ defined as the difference between prediction and reference values.
\label{fig:ml-prediction} 
}
\end{figure}

We first benchmark the trained ML model by comparing its predictions against the results from VMC simulations for all disorder configurations, including both the training and the validation datasets. As shown in Fig.~\ref{fig:ml-prediction}, for all four local quantities, the ML predictions agree reasonably well with the VMC calculations. More quantitatively, we plot the histograms of the prediction error defined as $\delta = A_{\rm ML} - A_{\rm VMC}$ in the insets of Fig.~\ref{fig:ml-prediction}. In all cases, the error is rather narrowly distributed with a MAE, given by the width of the distribution, of the order of less than one percent of the mean values.

It should be noted that the VMC dataset is itself noisy, since the quantities computed from VMC simulations are based on Monte Carlo samplings, for example, $n_{i} = \frac{1}{M} \sum_{i=1}^M \langle \Psi_i | \hat{n}_i | \Psi_i \rangle / \langle \Psi_i | \Psi_i \rangle$ for local electron occupation number, where $|\Phi_i \rangle$ is a trial wave function among the Markov chain samplings. The values from VMC thus depend on the number of samplings and other stochastic factors. Similarly, as the optimization of variational parameters, such as local energy renormalization $\Sigma_i$ and Gutzwiller parameter $g_i$, are based on derivatives which are computed stochastically, these quantities are also not without uncertainty. The randomness in the datasets thus partially contributes to the error $\delta$ in the ML predictions. Of course, the error due to VMC can be systematically reduced by increasing the number of Monte Carlo samplings.

We note in passing that for Hubbard-type models with electron-lattice coupling through the deformation potential, the ML prediction of the local electron number $n_i$ also provides the forces acting on the atomic displacements or lattice distortions. Our approach is thus an alternative to the more general Behler-Parrinello ML scheme~\cite{behler07}. These include both the Holstein and Jahn-Teller lattice models.  For example, the electron-phonon coupling in the Holstein model is described by $\hat{\mathcal{H}}_{\rm el\mbox{-}ph} = - g \sum_i \hat{n}_i Q_i$, where $Q_i$ denotes the amplitude of local structural distortion. The lattice degrees of freedom play an important role in the emergence of complex electronic textures driven by electron correlation in Hubbard-type models. In such applications, the lattice distortions $Q_i$ can be treated as classical dynamical variables, and the electronic forces acting on $Q_i$ can be obtained from Hellmann-Feynman theorem: $F_i = - \langle \partial \hat{\mathcal{H}}_{\rm el\mbox{-}ph} / \partial Q_i \rangle = g \langle \hat{n}_i \rangle$. The ML model developed here thus can also be combined with the Langevin dynamics method to enable large-scale dynamical simulations of Hubbard-Holstein model.

\begin{figure}
\includegraphics[width=0.99\columnwidth]{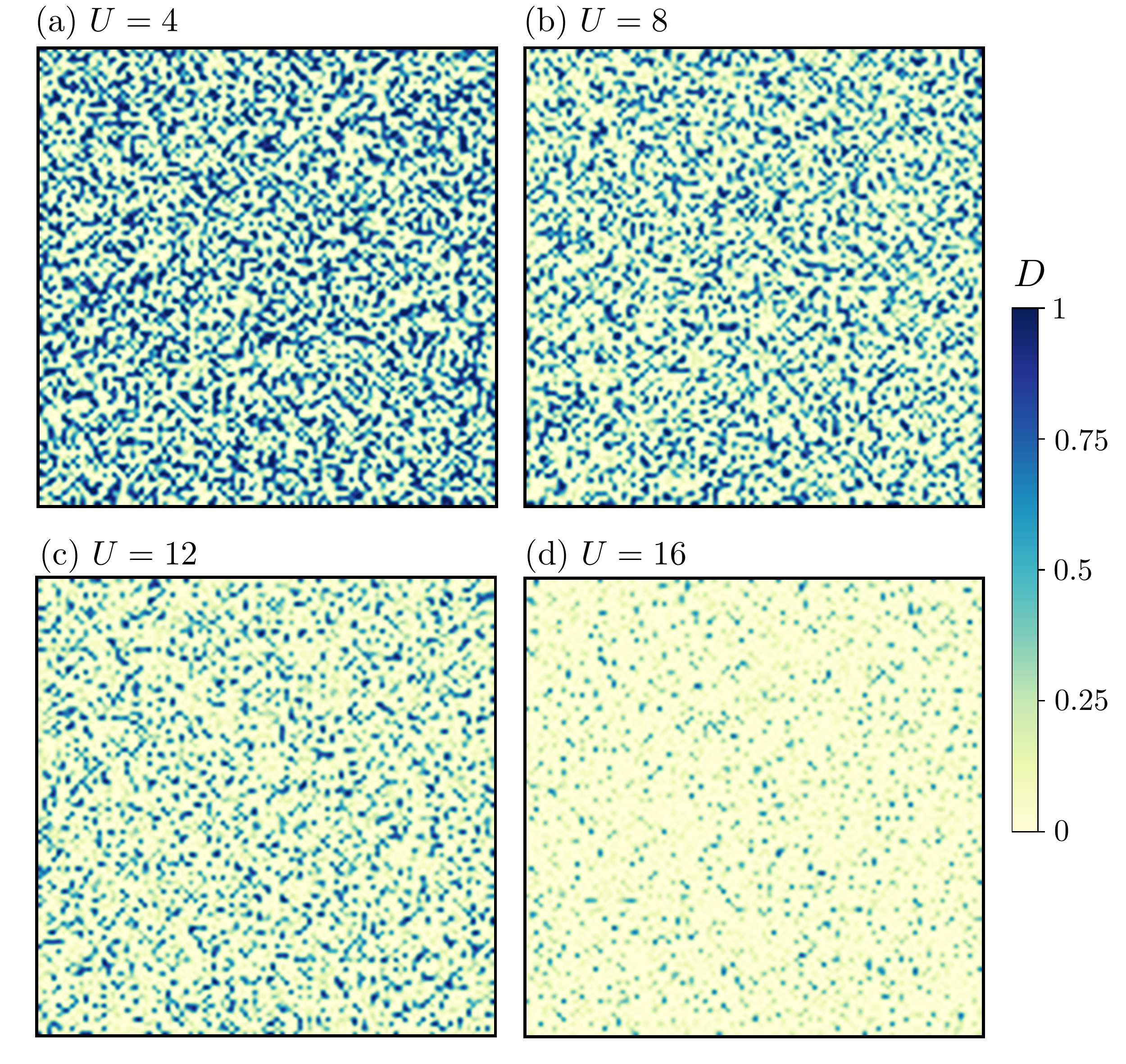}
\caption{Distributions of local double-occupancy $D_i$ obtained by applying the trained ML model to a large $100\times 100$ square lattice at various Hubbard parameters.
\label{fig:snapshots} 
}
\end{figure}

To demonstrate the scalability of our ML approach, we apply the ML model, trained from VMC solutions on a small $16\times 16$ lattice, to compute the real-space electronic properties of the AH model on a $100\times 100$ lattice. Fig.~\ref{fig:snapshots} shows the profiles of local double-occupancy $D_i$ at various Hubbard $U$ for some random realizations of a large disorder with $W = 18$. Based on the locality of electronic systems, the ML model only depends on on-site potentials of a fixed finite spatial region, e.g. determined by the cutoff radius $r_c$, independent of the system size~$N$. Consequently, the time complexity of ML method for computing local electronic properties scales {\em linearly} with~$N$. The efficiency is thus significantly improved compared with the polynomial scaling $\mathcal{O}(N^s)$ of VMC, where the exponent $s$ ranges from 3 to 6 depending on the specific optimization techniques~\cite{becca17}.

\begin{figure}[t]
\includegraphics[width=0.99\columnwidth]{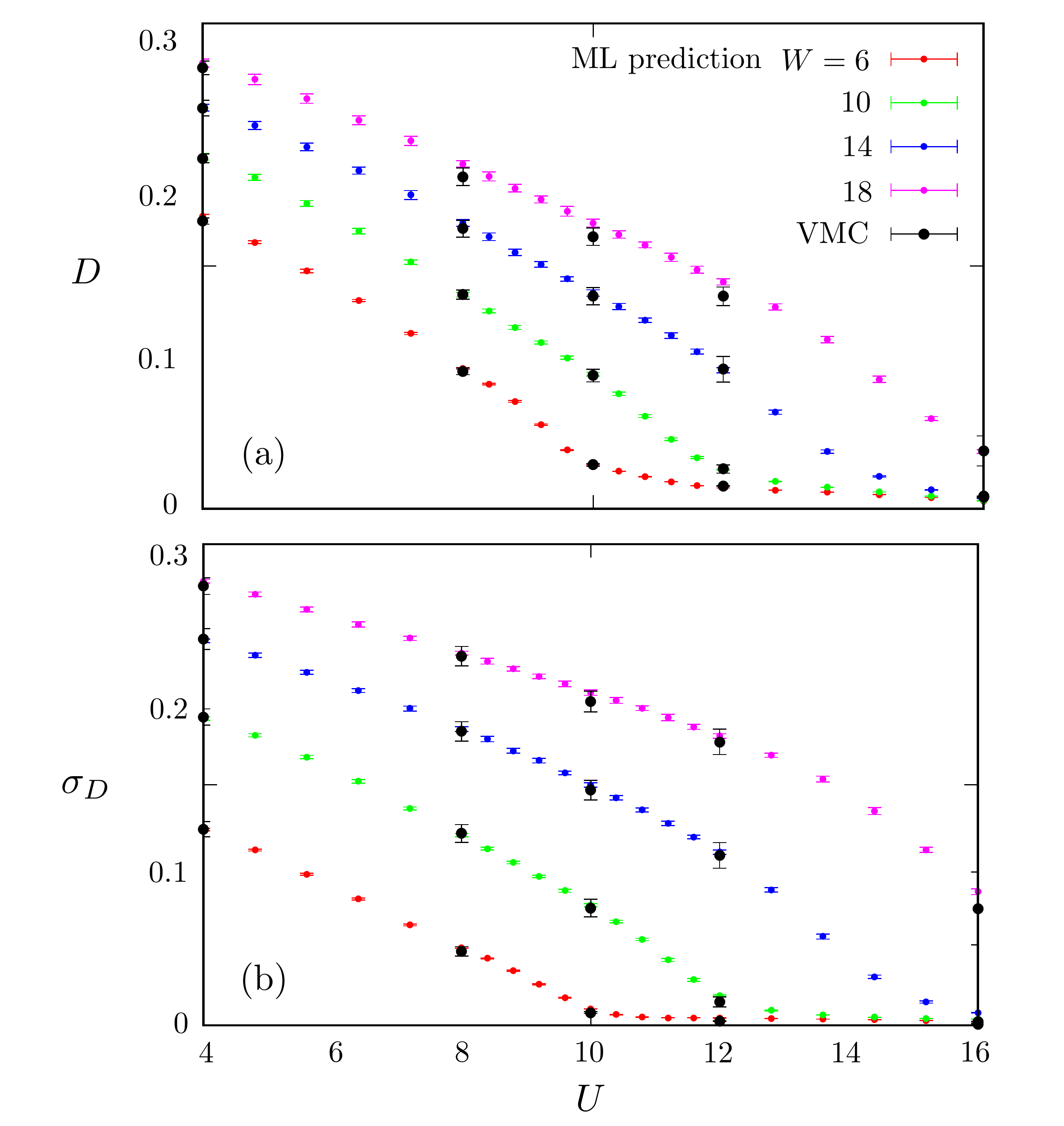}
\caption{ (a) The double occupancy $D$, averaged spatially over sites of the $100\times 100$ lattice and averaged over 50 independent realizations of disorder, as a function of Hubbard parameter~$U$. Panel (b) shows the spatial standard deviation of the double-occupancy $\sigma_D = [ \sum_{i=1}^N (D_i - \overline{D})^2 / N ]^{1/2}$, averaged over 50 different disorder realizations.   The black dots are the VMC data while the colored dots are predictions by the trained ML model.
\label{fig:dbl} 
}
\end{figure}

The results shown in Fig.~\ref{fig:snapshots} are also consistent with the picture of disorder screening discussed previously~\cite{andrade09,andrade10,tanaskovic03}. At small $U$, the strong disorder results in a highly inhomogeneous distribution of electron double occupation, as demonstrated in Fig.~\ref{fig:snapshots}(a). With increasing $U$, in addition to a reduced double occupancy due to Coulomb repulsion, the spatial inhomogeneity of $D_i$ is also reduced due to the renormalization $\Sigma_i$ of the on-site potential by electron correlation. By applying the ML model to independent realizations of disorder, Fig.~\ref{fig:dbl}(a) shows the double-occupancy $D$, averaged over all lattice sites and different disorder configurations, as a function of $U$ for different disorder strength $W$. Also shown for comparison is the VMC results on a smaller $16\times 16$ system. The ML predictions not only are consistent with the VMC calculations, but also exhibit a consistent trend towards Mott transition with increasing Hubbard parameter $U$. In the presence of strong disorder, our results show a continuous crossover from Anderson insulator to the Mott insulating phase~\cite{pezzoli10,villagran20}.

As shown in Fig.~\ref{fig:snapshots}, the rather inhomogeneous states indicate a rather wide distribution of the local double-occupancy, especially at small $U$. In addition to the mean value, our ML model also captures this spatial inhomogeneity of the electron state. To demonstrate this, we plot in Fig.~\ref{fig:dbl}(b) the spatial variation of the double-occupancy $\sigma_D = \bigl[ \frac{1}{N} \sum_i (D_i - \overline{D})^2 \bigr]^{1/2}$, averaged over several independent disorder realizations, as a function of~$U$. The amount of variation computed from ML predicted local double-occupancy agree very well with the VMC simulations. The ML model also consistently predicts the decrease of the dispersion $\sigma_D$ as $U$ is increased, which is indicative of the screening of disorder induced by strong electron correlation~\cite{andrade09,andrade10,tanaskovic03}. 

Finally, it is worth pointing out that the nontrivial $U$-dependence, which encapsulates the electron correlation effect, can be incorporated in the ML model by simply adding the Hubbard parameter $U$ at the input node; see Fig.~\ref{fig:ml-model}. The feasibility of this approach can be partly attributed to the smooth dependence of electronic properties on the Hubbard~$U$ in the presence of disorder. Indeed, as discussed in previous works, the first-order Mott transition is smeared by the strong disorder in 2D, leading to a continuous crossover to the Anderson-Mott insulator~\cite{pezzoli10,villagran20}. Moreover, even through a single input node, highly nonlinear dependence on $U$ can be achieved through the fully connected neurons with nonlinear activations.



\section{Summary and outlook}

\label{sec:summary}

To summarize, we present a comprehensive ML framework for the predictions of local electronic properties of disordered Hubbard models. By exploiting the universal approximation capability of neural networks, a ML model is developed to encode the complex dependence of local quantities, such as electron number and double occupancy, on the local environment. Based on group-theoretical method, a descriptor is proposed to represent the neighborhood random potentials with the lattice symmetry properly taken into account. We use the Anderson-Hubbard model as an example to demonstrate our ML framework. By training the NN with datasets from small-scale VMC simulations, we show that consistent results are obtained by applying the ML model to large systems with approximately $10^4$ lattice sites. 

Our approach emphasizes the scalability and transferability of the ML model, which are essential in order to achieve the goal of multi-scale modeling of correlated electron systems. The fact that most electronic structure methods and many-body techniques for solving strongly correlated models have a polynomial complexity $\mathcal{O}(N^\alpha)$ with $\alpha > 1$ significantly restricts the accessible size and time scales. On the other hand, as pointed out by W. Kohn, the locality nature of most electronic systems, namely, local physical observables are determined by the immediate environment, underpins the possibility of linear-scaling electronic structure methods. The ML scheme proposed in this work takes advantage of this feature to enable linear-scaling calculation for local electronic properties of correlated electron systems. 

It should be noted that similar approaches have been employed to develop ML force-field models for quantum MD simulations. However, in most prior works~\cite{behler07,bartok10,li15,botu17,li17,smith17,zhang18dp,behler16,deringer19,mcgibbon17,suwa19,mueller20,noe20,suwa19}, the ML models are derived from electronic structure methods that are based on self-consistent single-particle calculation, such as DFT or Gutzwiller approximation. Our work demonstrates that NN can successfully learn variational Monte Carlo simulation, which is a many-body method beyond effective single-particle or mean-field type approximations.

On the other hand, similar ML framework has also been applied to enable large-scale dynamical simulations of some correlated electron systems, including the double-exchange and Falicov-Kimball models~\cite{zhang20,zhang21,zhang21a}. These electronic models are characterized by dynamical classical degrees of freedom coupled to free electrons described by a tight-binding Hamiltonian. ML models are constructed for the generalized force fields acting on the dynamical classical fields. As there is no direct electron-electron interactions in these models, the electronic structure problems can be solved by exact diagonalization. For correlated electron systems with Hubbard-type interactions, sophisticated many-body methods are required for an accurate solution of the electronic structure problems. As mentioned above, our ML model can be readily combined with Langevin method to enable large-scale dynamical simulations of Hubbard-Holstein model.  Our work paves the way toward multi-scale dynamical modeling of strongly correlated systems such as Hubbard or $t$-$J$ models. 

\begin{acknowledgements}
This work was supported by the US Department of Energy Basic Energy Sciences under Award No. DE-SC0020330. Y.~H. Liu and T.~K. Lee are grateful for the support of the Taiwan Ministry of Science and Technology Grant MOST: 110-2112-M-110-018. The authors also acknowledge the support of  the Academia Sinica Grid-computing Center (ASGC) at Taiwan, as well as the Research Computing at the University of Virginia.
\end{acknowledgements}

\end{document}